\documentstyle[epsf,twoside,fleqn,espcrc2]{article}

\title{
 Phase structure of lattice QCD at finite temperature
 for 2+1 flavors of Kogut-Susskind quarks
\thanks{
{Presented by S.\ Kaya.
}}}
\author{JLQCD Collaboration : S. Aoki\address{Institute of Physics,
        University of Tsukuba, Tsukuba, Ibaraki 305-8571, Japan},
        M. Fukugita\address{Institute for Cosmic Ray Research,
        University of Tokyo, Tanashi, Tokyo 188-8502, Japan},
        S. Hashimoto\address{High Energy 
        Accelerator Research Organization (KEK), Tsukuba, Ibaraki
        305-0801, Japan}, K-I. Ishikawa\address{Department of Physics,
        Hiroshima University, Higashi-Hiroshima, Hiroshima 739-8526, Japan},
        N. Ishizuka$^{\rm a,}$\address{Center for Computational Physics,
        University of Tsukuba, Tsukuba, Ibaraki 305-8577, Japan},\\
        Y. Iwasaki$^{\rm a,e}$, K. Kanaya$^{\rm a,e}$, T. Kaneda$^{\rm
        a}$, S. Kaya$^{\rm c}$
        , Y. Kuramashi$^{\rm c}$, M. Okawa$^{\rm
        c}$, T. Onogi$^{\rm d}$,\\ S. Tominaga$^{\rm c}$, N. Tsutsui$^{\rm
        d}$, A. Ukawa$^{\rm a,e}$,
        N. Yamada$^{\rm d}$, T. Yoshi\'e$^{\rm a,e}$}

\begin{document}
\renewcommand{\textfraction}{0.1}
\renewcommand{\topfraction}{0.9}
\begin{abstract}
We report on a study of the finite-temperature chiral transition 
on an $N_t=4$ lattice for $2+1$ flavors of Kogut-Susskind quarks. 
We find the point of physical quark masses to lie 
in the region of crossover, in agreement with results of previous 
studies. Results of a detailed examination of the $m_{u,d}=m_s$ case 
indicate vanishing of the screening mass of $\sigma$ meson
at the end point of the first-order transition.
\end{abstract}

\maketitle
\setcounter{footnote}{0}

\section{Introduction}
An important issue in finite-temperature lattice QCD 
is the determination of the nature of the chiral phase transition
for a realistic spectrum of light up and down quarks and a heavier 
strange quark. 
Despite its importance, past studies of this ``2+1'' case have been few.
For the Kogut-Susskind action, 
all of them have been made around 1990\cite{Gavai,Kogut,Brown}.

It was found in these studies that the chiral phase transition changes 
from a first-order transition to a crossover as the strange quark 
mass $m_s$ increases beyond a critical value $m_s^c$ for a fixed 
degenerate up and down quark mass $m_{u,d}$, in agreement with 
predictions of an effective $\sigma$ model of QCD\cite{Wilczek,Gavin}.  
Results were also obtained\cite{Brown,Kogut} which indicate the physical 
point of quark masses to lie in the region of 
crossover on the $(m_{u,d}, m_s)$ plane. 
However, these results were based on simulations made at only a few sets 
of quark masses. 
Clearly a more extensive study is called for to have a full understanding 
of the phase structure in the 2-parameter space of $(m_{u,d}, m_s)$.  
Here we report first results from our recent effort toward this goal. 

An interesting suggestion from a $\sigma$ model analysis\cite{Gavin}
is that the second-order transition expected at the critical strange quark 
mass $m_s^c$ is in the Ising universality class, with the massless mode 
provided  by the $\sigma$ meson. 
A novel feature of our work is a study of the screening 
mass of $\sigma$ to examine this point. 

Our simulations are performed for the temporal lattice size $N_t=4$. 
An $8^3\times4$ lattice is employed to make a survey of the phase structure
varying $\beta$, $m_{u,d}$ and $m_s$. 
A detailed investigation is then made along the 
flavor $SU(3)$ symmetric 
line $m_{u,d}=m_s$ by another series of simulations 
on a $16\times 8^2 \times 4$ and a $16^3 \times 4$ lattice. 
For each parameter set, (1--2)$\times 10^3$ trajectories of unit length 
are generated by the hybrid R algorithm.

\begin{figure}[tb]
\leavevmode
\epsfxsize=7.0cm
\epsfbox{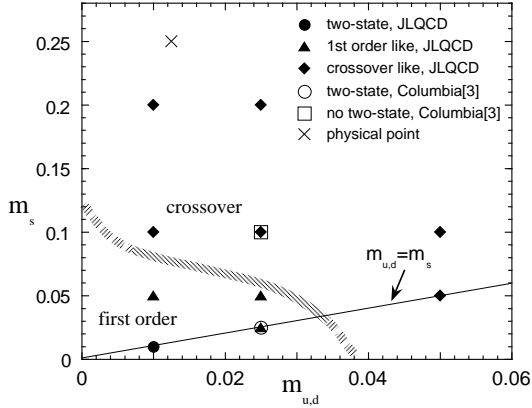}
\vspace{-1.1cm}
\caption{Phase diagram on $(m_{u,d},m_s)$ plane obtained on 
a $8^3\times 4$ lattice. Open symbols are results of 
Ref.~\protect\cite{Brown} on a $16^3\times 4$ lattice. Shaded line represents an
estimate of the critical line, and cross the physical point(see text). }
\label{fig:diag}
\vspace{-0.5cm}
\end{figure}

\section{Phase diagram on the $(m_{u,d},m_s)$ plane}
We show the result of our phase diagram analysis on an $8^3\times 4$ 
lattice in Fig.~\ref{fig:diag}.  
At $m_{u,d}=m_s=0.01$ a clear two-state signal is obtained by a comparison of 
runs with a hot and a cold start.
At triangle points, we find a very sharp change of observables
over a narrow range of $\beta$, suggestive of a 
first-order transition, while only a smooth 
crossover is seen at diamond points.

\begin{figure}[tb]
\vspace{-3mm}
\begin{center}
\leavevmode
\epsfxsize=7.0cm
\epsfbox{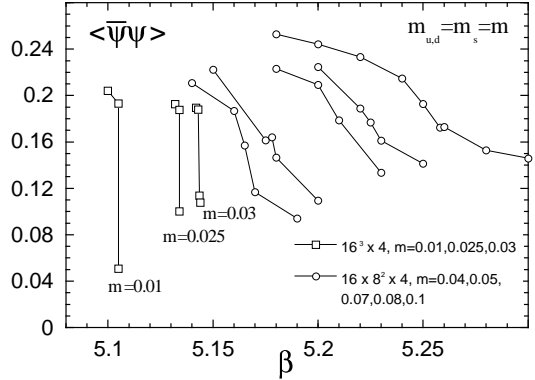}
\end{center}
\vspace{-1.5cm}
\caption{$\langle\overline{\psi}\psi\rangle$ as a 
function of $\beta$ for various quark masses along the line 
$m_{u,d}=m_s\equiv m$.}
\label{fig:fig2}
\vspace{-0.5cm}
\end{figure}

In Fig.~\ref{fig:fig2} we plot results for the chiral condensate along 
the line $m_{u,d}=m_s\equiv m$ on a $16^3\times 4$ lattice ($m\leq 0.03$) 
or a $16\times 8^2\times 4$ lattice ($m\geq 0.04$).  On these 
lattices a two-state signal is found, which is clear at $m=0.01$ and 0.025,  
but less so at $m=0.03$. The behavior above $m=0.04$, on the 
other hand, indicates a crossover. 

The results taken together suggest that the critical line marking the 
end point of the first-order transition runs above the triangle points 
at $m_{u,d}\leq 0.025$, and below the point $m_{u,d}=m_s=0.04$ in 
Fig.~\ref{fig:diag}, which is indicated by the shaded line. 
If we assume $T_c=150$~MeV for the critical temperature, the physical point 
is located at the cross in Fig.~\ref{fig:diag}, 
which is in the region of crossover.
Our results are in agreement with those of the Columbia 
group obtained at $m_{u,d}=0.025$ on a $16^3\times 4$ lattice\cite{Brown}.

\section{Results along the $m_{u,d}=m_s$ line}

We now discuss results along the line $m_{u,d}=m_s\equiv m$.  
Given our observation of a two-state signal at $m=0.01, 0.025$, one 
way to estimate the value of the critical mass $m^c$ is to 
extrapolate the gap of the 
chiral condensate $\Delta\langle\overline{\psi}\psi\rangle$ 
toward larger $m$ where it vanishes.
Employing the form
$\Delta\langle\overline{\psi}\psi\rangle\propto (m^c-m)^{1/2}$
predicted by the mean-field analysis of the $\sigma$ model, 
we find $m^c \simeq 0.034$. 
If we use a naive linear extrapolation, we obtain $m^c \simeq 0.049$. 
A similar value $m^c\approx 0.045$ was previously reported\cite{Gottlieb} 
by a linear extrapolation applied to old results\cite{Kogut,Brown}.

In the region of crossover $m>m^c$, we expect the peak height of 
the chiral susceptibility 
$\chi_m$ to develop a singular behavior 
$\chi_m\propto (m-m^c)^{-z}$ as $m\to m^c$. 
We calculate $\chi_m$ for $m \ge 0.04$ with the histogram 
reweighting method.  Assuming $m^c=0.034$, 
we fit the peak height to the form above. 
A reasonable fit with $\chi^2/df=1.05$ is obtained 
with the value of the exponent $z=0.67(3)$, which    
is comparable to the Ising value $z\simeq 0.79$ 
and the mean-field value 2/3. 

\begin{figure}[tb]
\vspace{-2mm}
\begin{center}
\leavevmode
\epsfxsize=7.0cm
\epsfbox{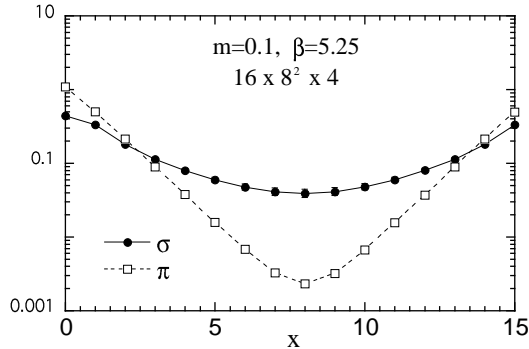}
\end{center}
\vspace{-1.5cm}
\caption{$\pi$ and $\sigma$ propagators for $m=0.1$ 
at the transition point $\beta$=5.25.}
\label{fig:fig3}
\vspace{-0.5cm}
\end{figure}

In order to examine the screening mass $M_{\sigma}$ of $\sigma$ meson,
we employ $U(1)$ random source and no gauge fixing to evaluate 
the two quark loop contribution of the $\sigma$ propagator. 
Good results are obtained for the full $\sigma$ propagator 
with this method as illustrated in Fig.~\ref{fig:fig3}.  

\begin{figure}[tb]
\vspace{-2mm}
\begin{center}
\leavevmode
\epsfxsize=7.0cm
\epsfbox{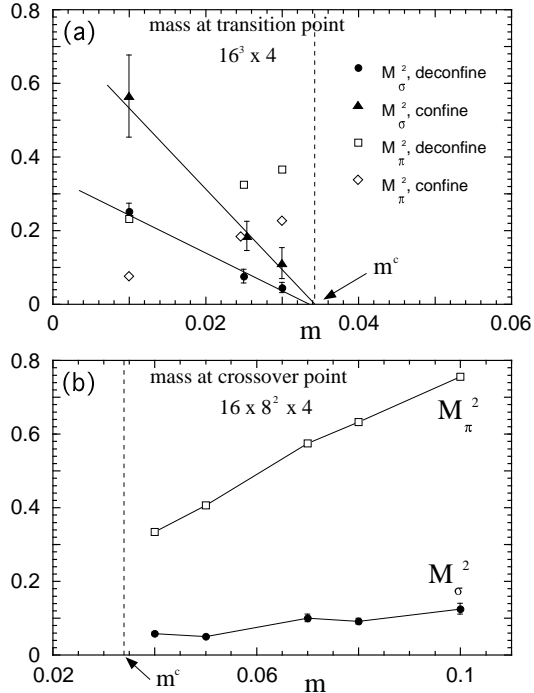}
\end{center}
\vspace{-1.5cm}
\caption{Screening masses $M_{\sigma}^2$ and $M_{\pi}^2$ as a function of $m$
for (a) $m\leq 0.03$ and (b) $m\geq 0.04$ at the transition point. 
In (a) results on the two sides of first-order transition are shown.}
\label{fig:fig4}
\vspace{-0.5cm}
\end{figure}

The quark mass dependence of $\pi$ and $\sigma$ screening masses 
for $m\leq 0.03$, where we find a first-order transition, 
is plotted in Fig.~\ref{fig:fig4}(a). 
We observe that  $M_{\sigma}^2$ 
decreases toward zero as $m$ increases toward the critical value,  
both on the confined  and the deconfined side of the transition,   
in contrast to $M_\pi^2$ which increases. 

Assuming a linear quark mass dependence,
$M_{\sigma}^2 \propto m^c-m$, predicted
by the mean-field analysis of the $\sigma$ model,
we obtain $m^c=0.034(3)$ in the confining
phase and $m^c=0.031(3)$ in the deconfining phase.
These values are consistent with each other, 
and are also in agreement with the estimate from a square root 
extrapolation 
of the gap of $\langle\overline{\psi}\psi\rangle$ discussed above.
These results indicate vanishing of the $\sigma$ mass at the critical 
quark mass as suggested by the $\sigma$ model\cite{Gavin}. 

Results for larger quark masses ($m\geq 0.04$), where a crossover 
behavior is seen, is shown in Fig.~\ref{fig:fig4}(b). 
While $M_\sigma^2$ decreases toward smaller values of $m$, 
the variation is too mild to attempt an independent estimate of $m^c$. 
An interesting point which requires clarification is that $M_\sigma^2$ 
stays considerably small compared to $M_\pi^2$ even for large quark masses. 

\section{Conclusions and future work}
Our study with the Kogut-Susskind action 
supports the previous conclusion with this action that there is 
no finite-temperature phase transition 
for three flavors of quarks with physical masses.
This means that a discrepancy with the conclusion from the Wilson
action\cite{Iwasaki} still remains.

We also find a strong indication that the screening mass 
of $\sigma$ vanishes at the end point of the first-order transition 
along the line $m_{u,d}=m_s$. 

We plan to extend analyses
carried out here to the $m_{u,d}\neq m_s$ case 
to further explore the real-world QCD chiral transition.

\vspace*{2mm}
This work is supported by the Supercomputer Project No.32 (FY1998)
of High Energy Accelerator Research Organization (KEK),
and also in part by the Grants-in-Aid of the Ministry of 
Education (Nos. 08640404, 09304029, 10640246, 10640248, 10740107, 10740125).
S.K. and S.T. are supported by the JSPS Research Fellowship.


\begin{thebibliography}{3}
\bibitem{Gavai} R.Gavai {\it et al.}, Phys.Rev.Lett.\ 58(1987) 2519.
\bibitem{Kogut} J. Kogut {\it et al.}, Nucl. Phys. B295 [FS21] (1988) 480;
	Phys. Lett. B263 (1991) 101.
\bibitem{Brown} F.Brown {\it et al.}, Phys.Rev.Lett.\ 65(1990) 2491.
\bibitem{Wilczek} R.\ Pisarski {\it et al.}, Phys.\ Rev.\ D29 (1984) 338.
\bibitem{Gavin} S.\ Gavin {\it et al.}, Phys. Rev. D49 (1994) 3079.
\bibitem{Gottlieb} S.\ Gottlieb, 
Nucl.\ Phys.\ B(Proc.\ Suppl.) 20 (1991) 247. 
\bibitem{Iwasaki} Y.\ Iwasaki {\it et al.}, Phys.\ Rev.\ D54 (1996) 7010.

\end{thebibliography}
\end{document}